\documentclass[english,aps,pra,reprint,noshowpacs,superscriptaddress]{revtex4-1}   
\usepackage[T1]{fontenc}	
\usepackage[latin9]{inputenc}	
\usepackage{geometry}		
\usepackage{graphicx}
\usepackage[above,below]{placeins}	
\usepackage{times}
\usepackage{hyperref}  
\hypersetup{colorlinks=true,urlcolor=blue,citecolor=blue,linkcolor=blue}   
\begin{document}
\title{Student perceptions of the value of out-of-class interactions: Attitudes vs. Practice}

\author{Justyna P. Zwolak}
\thanks{{\it JPZ currently at QuICS, UMD, College Park, MD 20742, USA and NIST, Gaithersburg, MD 20899, USA }}
\affiliation{Department of Teaching and Learning, Florida International University, Miami, FL 33199, USA}
\affiliation{STEM Transformation Institute, Florida International University, Miami, Florida 33199, USA}

\author{Remy Dou}
\affiliation{Department of Teaching and Learning, Florida International University, Miami, FL 33199, USA}
\affiliation{STEM Transformation Institute, Florida International University, Miami, Florida 33199, USA}

\author{Eric Brewe}
\affiliation{Physics Department, Drexel University, Philadelphia, PA 19104, USA}
\affiliation{School of Education, Drexel University, Philadelphia, PA 19104, USA}

\begin{abstract}
From industry to government to academia, attracting and retaining science, technology, engineering, and mathematics majors is recognized as a key element of the 21$^{\text{st}}$ century knowledge economy. The ability to retain students seems to be intimately tied with understanding their immersion into the academic and social system of an institution. For instance, it has been noted that insufficient interactions with peers can lead to a low commitment to the university and, ultimately, affect one's decision about whether to drop out. Since nearly half of first-time students who leave a university by the end of the freshman year never come back to college, the importance of understanding experiences in introductory courses as a means for improving students' persistence is particularly pronounced. We investigate students' experiences in introductory physics courses, focusing on their self-reported perception of the value of out-of-class collaborations. We find that, even though students consider the out-of-class collaborations to be important for success, it takes a relatively long time before they start practicing collaborative learning.
\end{abstract}

\maketitle

\section{Introduction}

The ability to retain students seems to be intimately tied with understanding their immersion into the academic and social system of an institution. Insufficient interactions with others, as well as a lack of compatibility with social values of the institution, lead to a low commitment to the university~\cite{Gainen95-BTS}. Ultimately, this affects students' decisions about dropping out of a major or school altogether. A high degree of integration into the university, on the other hand, leads to greater commitment to the institution and to completing college~\cite{Tinto75-DHE}. Understanding the effects of immersion into the academic and social system of a university on students' persistence is crucial for improving their educational experience. For students who just started education and have not yet formed connections in the community, particularly those who commute to college, the classroom might be the only place where connecting with others happens. Thus, the importance of the classroom experience in introductory courses as a means for improving student persistence should not be underestimated.

As part of an ongoing effort to increase success rates of traditionally underrepresented students at Florida International University (FIU), we conducted a study that investigates the effect of students' social and academic integration on persistence. Using a social network analysis approach we identified patterns of interactions in an active engagement introductory physics classroom that improve the odds of persistence. In particular, we found that the number of peers that a given individual reached out to during class-time (i.e., {\it outdegree}) and the overall embeddedness within the in-class network of self-reported interactions (i.e., {\it closeness}) by the end of the Fall semester were positively correlated with the odds of continuing into the second course in the sequence~\cite{Zwolak17-IIP}.

Here, we take the analysis of students' integration one step further and investigate how students perceive the value of collaborative work. In particular, we want to determine whether students' attitudes toward working with peers change over the course of the Modeling Instructions (MI) introductory physics sequence. We also look into how their attitudes translate into actual behaviors and whether there are differences when group work is required (e.g., group lab reports, group exam) or just encouraged (e.g., homework assignments, study groups). In a group-work-based curriculum, such as MI, collaborative work is not only explicitly promoted, but actually incorporated into the curriculum. Thus, the MI classroom represents a distinct environment for examining how students' interactions relate to out-of-class academic activity.

\section{Methodology}\label{sec:method}

\subsection{Students' Experiences Questionnaire}
To better understand students' experiences in the MI courses, we designed a short Students' Experiences Questionnaire (SEQ) that was sent out to students enrolled in the MI courses after the final exam. SEQ sent in Fall 2016 included 13 questions about, among other things, students' initial expectations for the MI course, time they had to spend studying outside of class, and their attitudes toward learning with peers outside of class. In Spring 2017 the number of questions was reduced to 10. Additionally, to ease the completion of the SEQ, we converted all open-ended questions into multiple-choice format based on answers from Fall 2016. To do so, we classified open-ended Fall 2016 responses into categories representing similar meaning, following the conventional content analysis approach~\cite{Hsieh05-QCA}. We started by combining all answers to a particular question into a single document that was then read multiple times to derive codes. In stage one, we highlighted the phrases from the responses that were most relevant to the question being asked. In stage two, we looked at these key phrases to identify emergent categories. In stage three, all leftover answers were either assigned to one of the pre-existing categories or used to form new ones, depending on how the answers were linked to the original set of categories. The inter-rater reliability test was a part of stage four.

Qualtrics was used to manage and distribute the questionnaire both times. Email addresses for all students were retrieved from FIU's electronic record system. Prior to attempting the questionnaire, students were informed that their participation is voluntary and that their answers will not affect their grades. Once started, students had four hours to complete the questionnaire. After that time, responses with at least $75\%$ completion were recorded. Students who did not take the survey or did not finish at least $75\%$ of the questionnaire were sent electronic reminders to complete. The reminders were sent twice and the survey remained open for two weeks.
 
\subsection{Survey administration}
The SEQ was distributed twice: in the Fall 2016, after the MI mechanics (MI-M) course, and then in the Spring 2017, after the MI electricity and magnetism (MI-EM) course. There were two section of MI-M, taken by a total $126$ students ($73$ in section $S_A$ and $53$ in section $S_B$), and two sections of MI-EM, taken by a total of $127$ students ($72$ and $55$ in sections $S_A$ and $S_B$, respectively). There were two instructors teaching the course, each teaching one section in the Fall and one in the following Spring. Students enrolled in MI-M could either take the second course in the series with the same instructor as the first course, change instructors, or switch to a traditional lecture-based course. Students enrolled in MI-EM included both students who took MI-M and students coming from traditionally taught mechanics course. All students are expected to complete some format of introductory-level mechanics course prior to enrolling in MI-EM. 

In Fall 2016, some of the SEQ questions were personalized based on responses to our social network (SN) survey (see~\cite{Zwolak17-IIP} for details about the SN survey). In particular, phrasing of some of the SEQ questions depended on the frequency of responses to the SN survey. As a consequence, students who completed the SN survey less than two times were excluded from the data collection ($7$ students from section $S_A$ and $2$ students from section $S_B$~\cite{Note1}; $N{=}\,117$). In Spring 2017, all questions on the SEQ were simplified, regardless of the frequency of answers to the SN survey, and thus the SEQ was sent to all students enrolled in MI-EM ($N{=}\,127$).

We received $80$ responses in Fall 2016 ($43$ from section $S_A$, $37$ from section $S_B$) and $82$ in Spring 2017 ($47$ from section $S_A$, $35$ from section $S_B$). One student from Spring 2017 opted out from the study. The response rates were fairly comparable between semesters and sections ($S_A\colon\,65\%$, $S_B\colon\,73\%$ in Fall 2016 and  $S_A\colon\,67\%$, $S_B\colon\,64\%$ in Spring 2017). Finally, $47$ students responded to the SEQ in both semesters.
	
\section{Analysis and results}

\begin{figure}[b]
\includegraphics[width=1.0\linewidth]{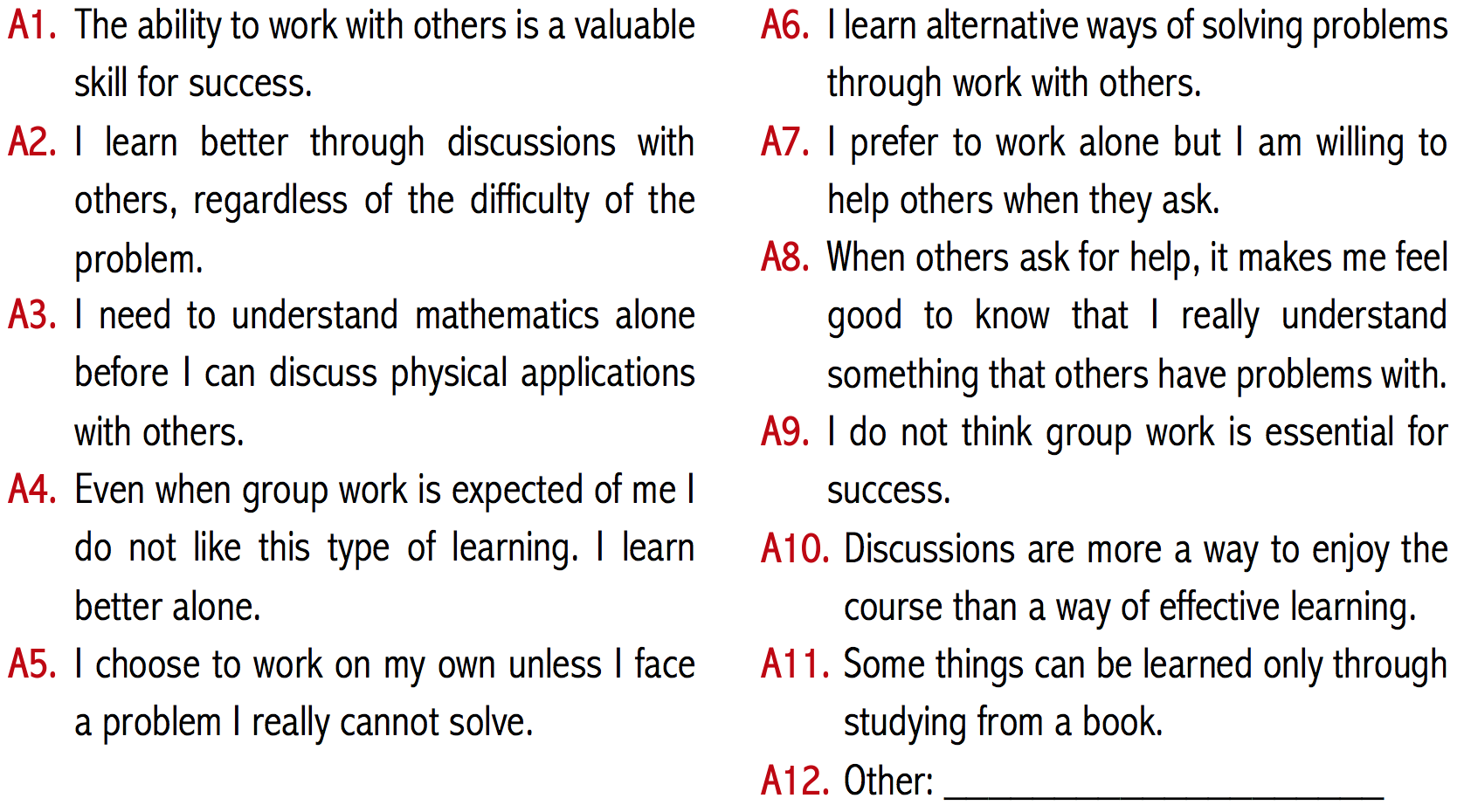}
\caption{Statements describing attitudes toward learning with peers that were presented to students as part of the SEQ survey.}\label{fig:attitudesANS}
\end{figure}

For this study, we are interested in students' attitudes toward collaborative learning and in how these attitudes translate into and reflect practice through actual collaborative work outside of class. To gain insight into students' attitudes, we asked them the following question: 
\begin{center}
\fbox{\parbox{0.9\linewidth}{Which of the following statements best describes your attitudes toward learning with peers (select all you agree with).}}
\end{center}
We then presented them with a list of options to choose from, as shown in Fig.~\ref{fig:attitudesANS}. The ``Other'' option was selected by only three individuals in Fall 2016 (none in Spring 2017). Two out of the three comments given as a part of this choice paraphrased existing choices, and the third one was a comment on how the course could be restructured. Because of the overlap in the first two comments and the lack of relevance of the third, we excluded the ``Other'' option from our analysis. 

\begin{figure}[t]
\includegraphics[width=1.0\linewidth]{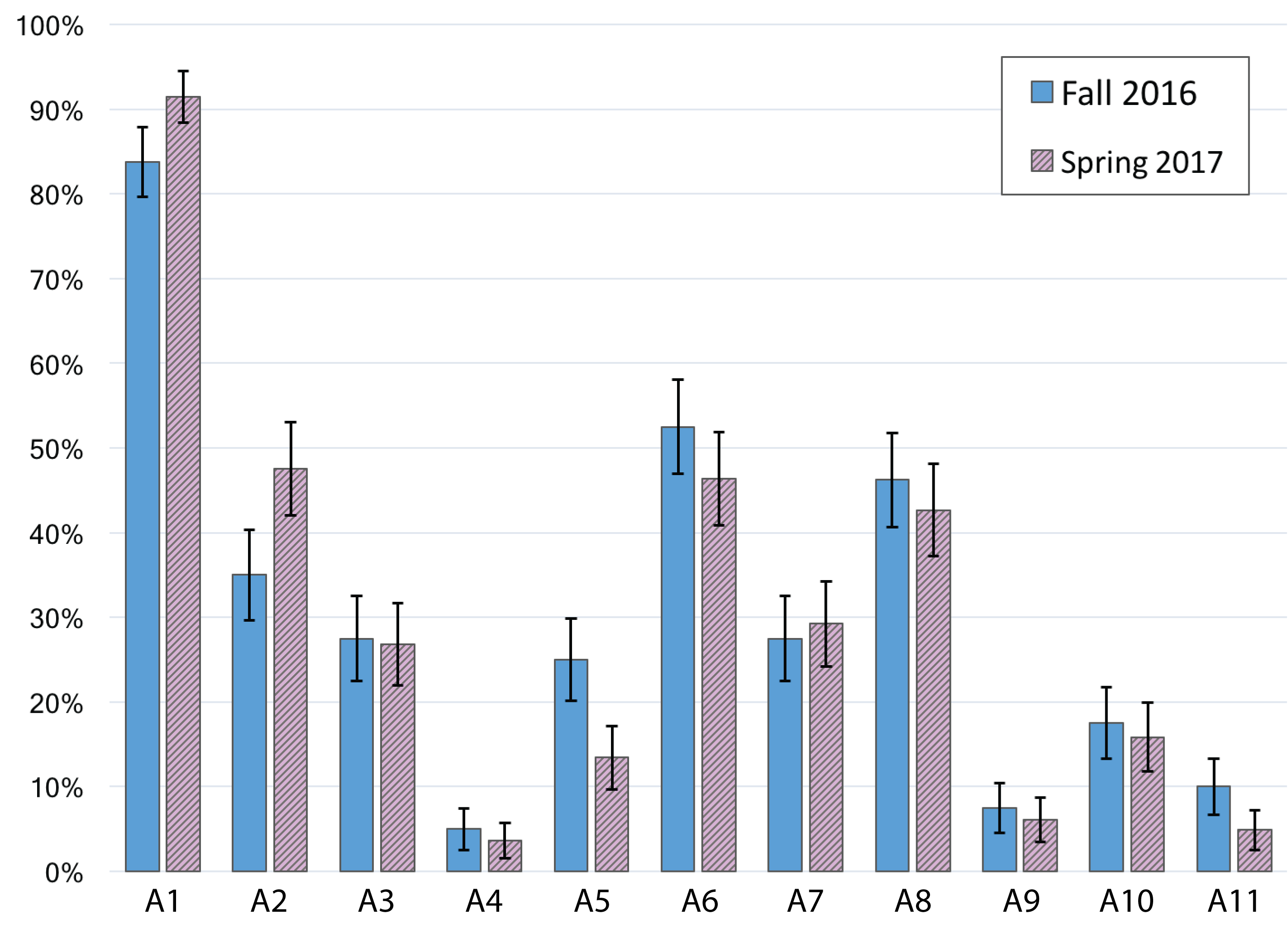}
\caption{Students' self-reported attitudes toward out-of-class collaborations: Fall 2016 (solid blue, $N{=}\,80$) and Spring 2017 (hatched pink, $N{=}\,82$). Responses A1 -- A11 are in Fig.~\ref{fig:attitudesANS}. The error bars represent the standard error of the sample estimates.}\label{fig:attitudesFvsS}
\end{figure}

Figure~\ref{fig:attitudesFvsS} shows responses to this question from Fall 2016 (solid blue) and Spring 2017 (hatched pink). We find that responses between the two semesters do not differ significantly (as verified by the two-sample test of proportions \cite{Note2}). In other words, we find no variation in attitudes towards group work between students taking introductory mechanics and introductory electricity and magnetism courses.

\begin{figure}[t]
\includegraphics[width=1.0\linewidth]{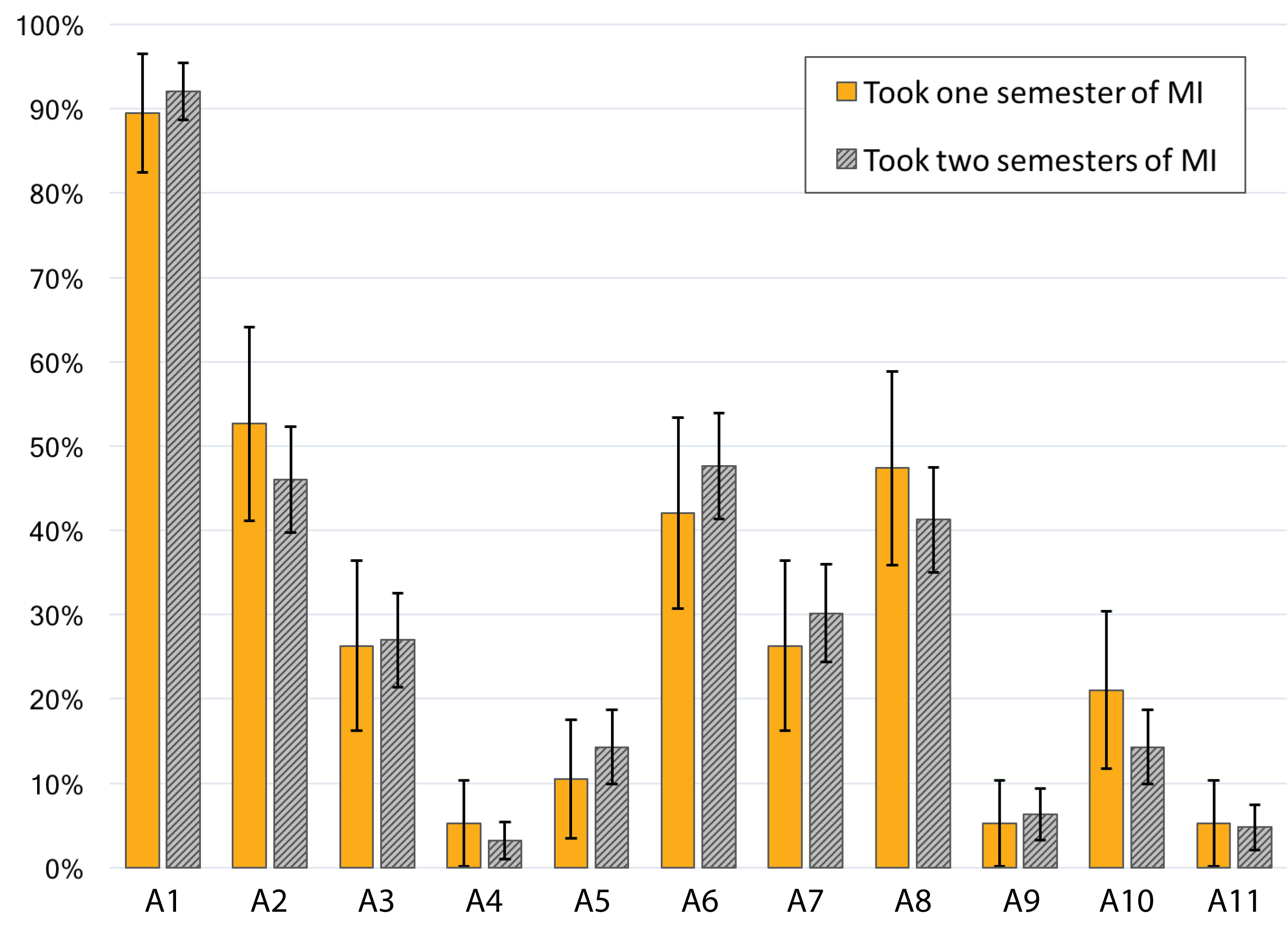}
\caption{Students' self-reported attitudes toward out-of-class collaborations: students who took one semester of MI (solid orange, $N{=}\,19$) and students who took two semesters of MI (hatched gray, $N{=}\,63$). Responses A1 -- A11 are in Fig.~\ref{fig:attitudesANS}. The error bars represent the standard error of the sample estimates.}\label{fig:attitudesS17}
\end{figure}

Since more than $75\%$ of respondents from Spring 2017 took two consecutive semesters of MI, it is natural to ask whether longer exposure to the active-engagement environment of MI may lead to a shift in attitudes toward working with peers. To determine whether attitudes of students who had prior experience with MI curriculum differ from those who were taking the MI course for the first time, we divide the Spring 2017 data into two groups: (1) students who took only one semester of MI (i.e., MI-EM), and (2) students who took two semesters of MI (i.e., MI-M and MI-EM). As can be seen in Fig.~\ref{fig:attitudesS17}, there are no significant differences in responses between the two groups. Interestingly, the overall pattern of responses is similar between Fig.~\ref{fig:attitudesFvsS} and Fig.~\ref{fig:attitudesS17}.

\begin{figure}[b]
\includegraphics[width=1.0\linewidth]{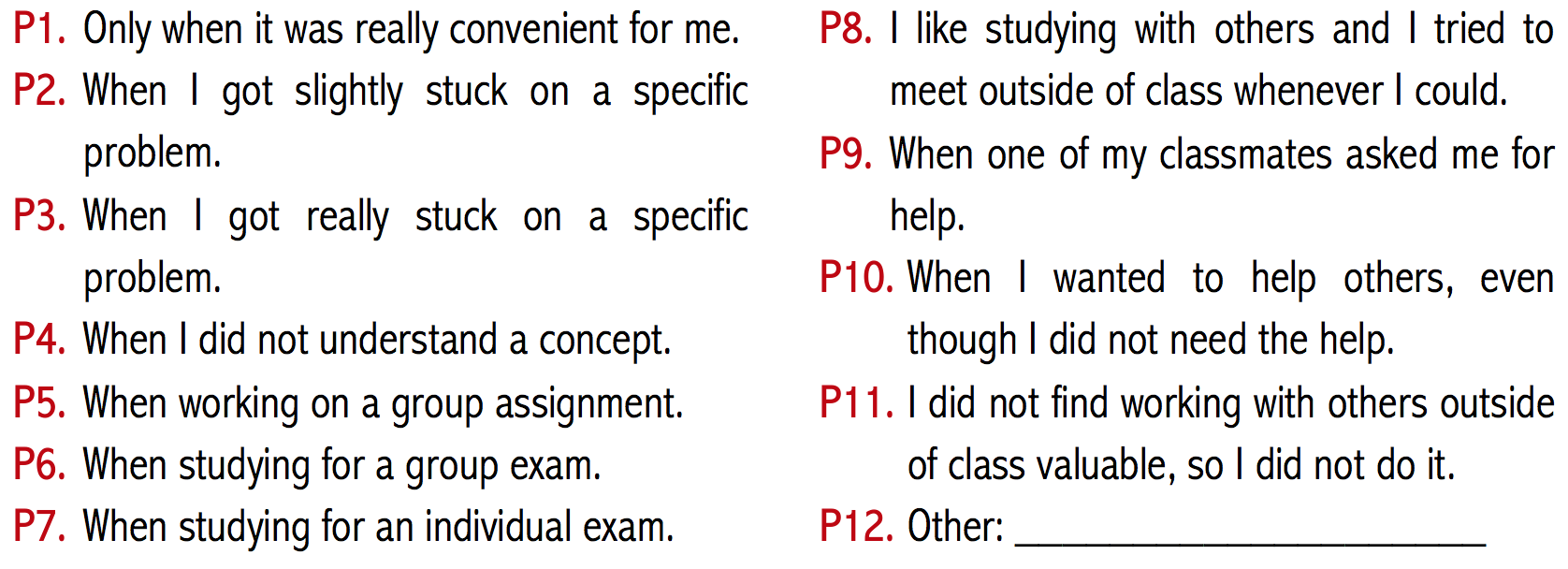}
\caption{Options that students could choose from to report collaborative work outside of class.}\label{fig:behavANS}
\end{figure}

\begin{figure}[t]
\includegraphics[width=1.0\linewidth]{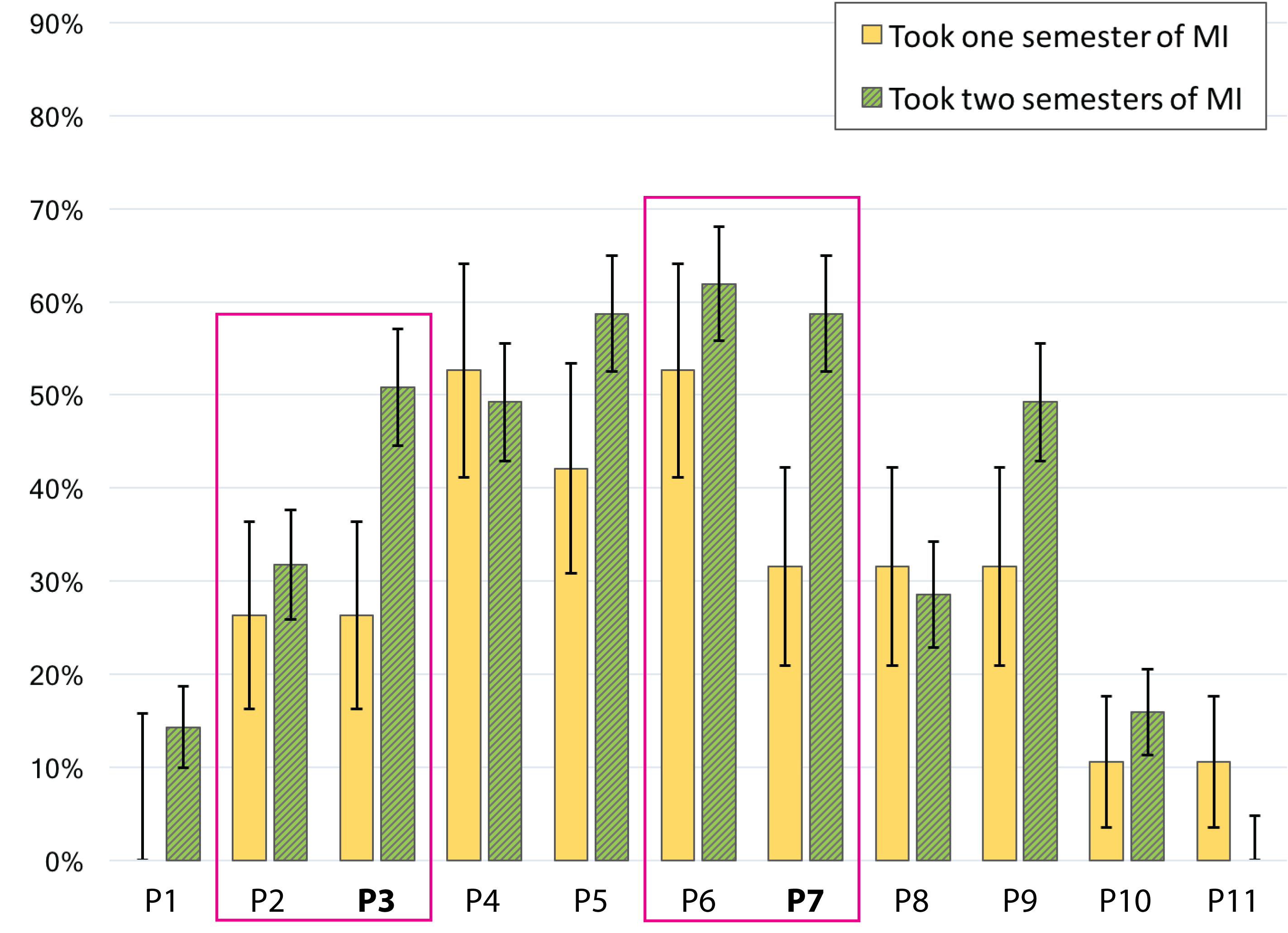}
\caption{Students' self-reported out-of-class collaborations in practice: students who took one semester of MI (solid yellow, $N{=}\,19$) and students who took two semesters of MI (hatched green, $N{=}\,63$). For categories P1 -- P11 see Fig.~\ref{fig:behavANS}. The error bars represent standard error of the sample estimates.}\label{fig:behav}
\end{figure}

Finding no differences in attitudes toward group work, regardless of prior experience with MI or lack thereof, we decided to investigate whether analogous situation occurs for students' reported collaborative work with others. To do so, we compare students' answers on a question about situations in which they did take time to work with peers:
\begin{center}
\fbox{\parbox{0.9\linewidth}{In what situations did you take the time to work on physics {\bf with others outside of class} (e.g., meeting right before/after class, meeting on campus, chatting on WhatsApp, using Google Docs, etc.)? (select all that apply)}}
\end{center}
Options that students could choose from are shown in Fig.~\ref{fig:behavANS}. We find that for 8 out of the 11 choices responses from students with no prior experience with MI present less favorable approaches toward group work than from students who took MI-M, with two options exhibiting statistically significant differences (see Fig.~\ref{fig:behav}). In particular, we find that although students in both groups report taking time to work with others when they are slightly stuck on a particular problem (P2; $26\%$ for group (1), $32\%$ for group (2)), students who took MI-M prior to taking MI-EM are much more likely to reach out for help when the problem is more challenging and they are ``really stuck'' (P3; $26\%$ and $51\%$ for group (1) and (2), respectively). We find a similar trend in how an approach to studying with others before exams. Students who took MI-M were nearly twice as likely to study with peers when preparing for {\it individual} exams than students with no prior MI experience (P7; $59\%$ and $32\%$, respectively). This held true even though an approach to working with peers when preparing for the {\it group} exam was comparable between both groups (P6; $53\%$ and $62\%$ for group (1) and (2), respectively). 

\section{Discussion}

Prior studies found that active engagement courses lead to building better networks of peer-to-peer connections~\cite{Brewe10-PLC}. Embeddedness in these networks have been shown to be a good predictor of persistence and performance~\cite{Zwolak17-IIP, Williams17-EII}. This aligns with prior research in introductory physics courses that indicates a clear learning advantage for students in large-enrollment courses taught using pedagogies that promote peer-to-peer interactions~\cite{Deslauriers11-ILP}. In other words, students who leverage the opportunities to work with others may find improved success rates in the so-called introductory ``gatekeeper'' STEM courses.
 
We investigate both students' self-reported perception of the values of out-of-class collaborations in introductory physics courses, and their collaborative behavior. We find that, even though students consider the out-of-class collaborations to be important for success, their willingness to actually engage with peers seems to depend on how much experience with the active engagement curriculum they have. In particular, students' beliefs about the value of collaborative activities with peers do not differ from one semester to the next (see Fig.~\ref{fig:attitudesFvsS}) and remains fairly stable regardless of whether prior experience with the MI curriculum is taken into account (i.e., taking only MI-EM versus taking both MI-M and MI-EM, see Fig.~\ref{fig:attitudesS17}).  Nevertheless, students who had prior experience with MI (i.e., those enrolled in MI-M), in general report greater out-of-class collaborative participation during the second course in the sequence (see Fig.~\ref{fig:behav}). In some cases, their participation nearly doubles. 
 
Interestingly, we find significant discrepancy between students' approach to exam preparation -- while there is no significant difference for the {\it group} exam (P6), for the {\it individual} exams (P7) students who took two semesters of MI seem to see the benefits of study groups more than those who have only one semester of experience. Similarly, when slightly stuck on a problem (P2), students from both groups reported working with peers with comparable frequency but, when the problems were more challenging (P3), students with two semesters of MI were much more likely to reach out to peers. Students with no prior MI experience reported working on challenging problems with peers just as often as they reported for somewhat difficult problems. For the group with two semesters of MI experience, the reported collaborations nearly doubled for challenging problems when compared with somewhat difficult ones. Finally, we find that the overall trend is more favorable for group work among students with two semesters of MI. 
 
Most importantly, even though students perceive teamwork as a valuable skill for success ($92\%$ for Spring 2017), only about half of them reported working with peers, even when the teamwork was expected ($60\%$ of students reported studying together before the {\it group} exam) or explicitly requested ($55\%$ reported working with peers on group assignments). Not even a third of students (about $30\%$) said that they actually like studying with others.
 
To conclude, it seems like it takes quite a long time for students to begin to actively collaborate, even with regular exposure to peer-to-peer learning. Those who were experiencing MI curriculum for the first time worked with peers when they had to (e.g., before {\it group} exam) but not so much in other situations, even though it would be just as beneficial for them (e.g., before {\it individual} exams). One reason for this might be that students are used to the traditional style of teaching where information is passed on to them during the lecture or acquired from books. Learning with and from peers is a very different setup that requires a significant shift in mindset. Our analysis suggests that it might take basically an entire semester of MI before students develop the practice of reaching out to peers. This hypothesis is further supported by interview data, although a further analysis that takes into account the effect of knowing peers from the previous course is needed~\cite{in.prep}. Our analysis shows that there is, however, a discrepancy between students' attitudes toward and practicing of collaborative learning. In order to help with the transition from traditional teaching to peer-to-peer learning this discrepancy should be taken into account.

\acknowledgments{Supported by NSF PHY 1344247.}


\end{document}